\documentclass{article}
\usepackage{arxiv}
\usepackage{graphicx}%
\usepackage{multirow}%
\usepackage{amsmath,amssymb,amsfonts}%
\usepackage{amsthm}%
\usepackage{mathrsfs}%

\usepackage[utf8]{inputenc} 
\DeclareUnicodeCharacter{2500}{--} 
\usepackage[T1]{fontenc}    
\usepackage{hyperref}       
\usepackage{url}            
\usepackage{booktabs}       
\usepackage{amsfonts}       
\usepackage{nicefrac}       
\usepackage{microtype}      
\usepackage{lipsum}
\usepackage{graphicx}
\graphicspath{ {./images/} }

\usepackage{textcomp}%
\usepackage{booktabs}%
\usepackage{algorithm}%
\usepackage{algorithmicx}%
\usepackage{algpseudocode}%
\usepackage{listings}%
\usepackage{makecell}
\usepackage{array}
\usepackage{float}

\usepackage{comment}
\usepackage[table]{xcolor}
\definecolor{coral}{RGB}{255,127, 80}

\usepackage{xr-hyper}  %
\usepackage{hyperref} 
\usepackage{threeparttable,booktabs}

\title{
Flow matching for reaction pathway generation
}

\author{
 Ping Tuo \\
  Bakar Institute of Digital Materials for the Planet\\
  University of California, Berkeley\\
  Berkeley, 94720, CA, United States \\
  \texttt{tuoping@berkeley.edu} \\
   \And
 Jiale Chen \\
  Institute of Science and Technology Austria \\
  Am Campus 1, 3400, Klosterneuburg, Austria \\
  \And
 Ju Li \\
  Department of Nuclear Science and Engineering and Department of Materials Science and Engineering\\
  Massachusetts Institute of Technology\\
  Cambridge, 02139, MA, United States \\
  \texttt{liju@mit.edu} \\
}

\renewcommand{\today}{}

\begin{document}
\maketitle
\begin{abstract}
Elucidating reaction mechanisms hinges on efficiently generating transition states (TSs), products, and complete reaction networks. Recent generative models, such as diffusion models for TS sampling and sequence-based architectures for product generation, offer faster alternatives to quantum-chemistry searches. But diffusion models remain constrained by their stochastic differential equation (SDE) dynamics, which suffer from inefficiency and limited controllability. We show that flow matching, a deterministic ordinary differential (ODE) formulation, can replace SDE-based diffusion for molecular and reaction generation. We introduce MolGEN, a conditional flow-matching framework that learns an optimal transport path to transport Gaussian priors to target chemical distributions. On benchmarks used by TSDiff and OA-ReactDiff, MolGEN surpasses TS geometry accuracy and barrier-height prediction while reducing sampling to sub-second inference. MolGEN also supports open-ended product generation with competitive top-k accuracy and avoids mass/electron-balance violations common to sequence models. In a realistic test on the $\gamma$-ketohydroperoxide decomposition network, MolGEN yields higher fractions of valid and intended TSs with markedly fewer quantum-chemistry evaluations than string-based baselines. These results demonstrate that deterministic flow matching provides a unified, accurate, and computationally efficient foundation for molecular generative modeling, signaling that flow matching is the future for molecular generation across chemistry.
\end{abstract}


\section{Introduction}
\label{sec1}
Predicting the outcomes of chemical reactions from given reactants and conditions remains a central yet formidable challenge in chemistry. Expert chemists typically rely on mechanistic reasoning, often visualized through ``arrow-pushing" diagrams to anticipate reaction pathways. However, identifying plausible mechanisms purely through intuition is frequently insufficient, motivating the development of computational and quantitative approaches for mechanistic exploration.

There have been several classes of automated methods for generating reaction networks from specified reactants using quantum chemical calculations~\cite{dewyer2018methods}. Template-based approaches encode known elementary reaction rules to construct networks of potential intermediates and products~\cite{rappoport2014complex}. Physics-based methods identify transition states to assess kinetic feasibility, followed by intrinsic reaction coordinate analyses to trace reaction pathways~\cite{ohno2004scaled,maeda2005global,maeda2011finding,maeda2013systematic,garay2022new,bhoorasingh2015transition,martinez2015automated}. Graph-based algorithms represent molecules as graphs and recursively enumerate intermediates accessible via single-step transformations~\cite{dewyer2017finding,yang2017automatic}, subsequently linking reactants and products through techniques such as the growing string and nudged elastic band methods.
While these strategies provide powerful frameworks for elucidating mechanisms and discovering novel pathways, they remain computationally demanding.
However, constructing even a moderately sized reaction network can require thousands of elementary steps, demanding millions of DFT single-point evaluations~\cite{von2020exploring}. This computational bottleneck has spurred the development of machine learning approaches capable of modeling complex chemical transformations with greater efficiency.

Diffusion models have emerged as state-of-the-art approaches for molecular generation, wherein a neural network learns the underlying probability distribution and generates new configurations by modeling a reverse-time stochastic diffusion process. This process gradually transforms a simple prior distribution (e.g., Gaussian noise) into a complex data distribution. 
Numerous diffusion frameworks have been developed for central problems in physics and chemistry~\cite{xie2021crystal,zeni2023mattergen,jiao2023crystal}. 
In mechanistic chemistry, diffusion-based approaches such as TSDiff~\cite{kim2024diffusion} and OA-ReactDiff~\cite{duan2023accurate} have been introduced to generate the transition states of each elementary reaction in a reaction network, replacing the costly nudged elastic band or string methods. However, their reliance on stochastic differential equations (SDEs) introduces limitations familiar to all diffusion approaches: stochastic noise complicates training, slows sampling, and makes conditional generation difficult to control.

Flow matching offers a deterministic alternative. Instead of learning a stochastic reverse diffusion process, a flow-matching model learns the velocity field of an ordinary differential equation (ODE) that continuously transports a prior distribution to a target distribution. This replaces the noisy SDE formulation of diffusion models with a smooth, analytically defined flow, yielding stable training, precise trajectory control, and dramatically faster sampling. Conceptually, flow matching unifies the family of SDE-based diffusion models, rectified flows~\cite{liu2022flow}, and consistency models~\cite{song2023consistency} under a single deterministic transport framework.

Here, we show that flow matching can entirely replace diffusion-based formulations in reaction mechanism generation. We introduce MolGEN, a flow-matching framework that learns to generate transition states (TSs), reaction products, and multi-step reaction networks directly from benchmark datasets previously used by TSDiff and OA-ReactDiff.
MolGEN halves the TS geometry prediction error relative to diffusion models, while reducing sampling time to sub-second. Remarkably, it also uncovers alternative transition states with lower barrier heights than those reported in the reference data, underscoring its potential to reveal previously inaccessible mechanistic pathways.
In Section~\ref{sec:gen-p}, we assess MolGEN on reaction product generation from given reactants, where it attains top-$k$ accuracies on par with the most advanced sequence-based approaches. Together, these results demonstrate that a deterministic flow formulation can provide a self-contained, efficient, and scalable route to the automated rollout of complex reaction networks.
In Section~\ref{sec:rng}, we apply this framework to the decomposition network of $\gamma$-ketohydroperoxide (KHP). In this realistic benchmark, MolGEN delivers a substantial computational efficiency gain over conventional string-based methods, while maintaining high mechanistic fidelity.
The combination of high-quality generation, minimal computational cost, and broad task generality establishes flow matching as a powerful and generalizable paradigm for molecular and reaction generation.

\section{A general-purpose conditional flow matching design}

A flow matching model learns a transformation from an initial distribution, typically a standard Gaussian, to a target distribution. This transformation is defined by an optimal transport path connecting the source and target distributions. A neural network is trained to approximate the corresponding velocity field that governs the evolution of the interpolant along this path.

\begin{figure}
    \centering
    \includegraphics[width=0.55\linewidth]{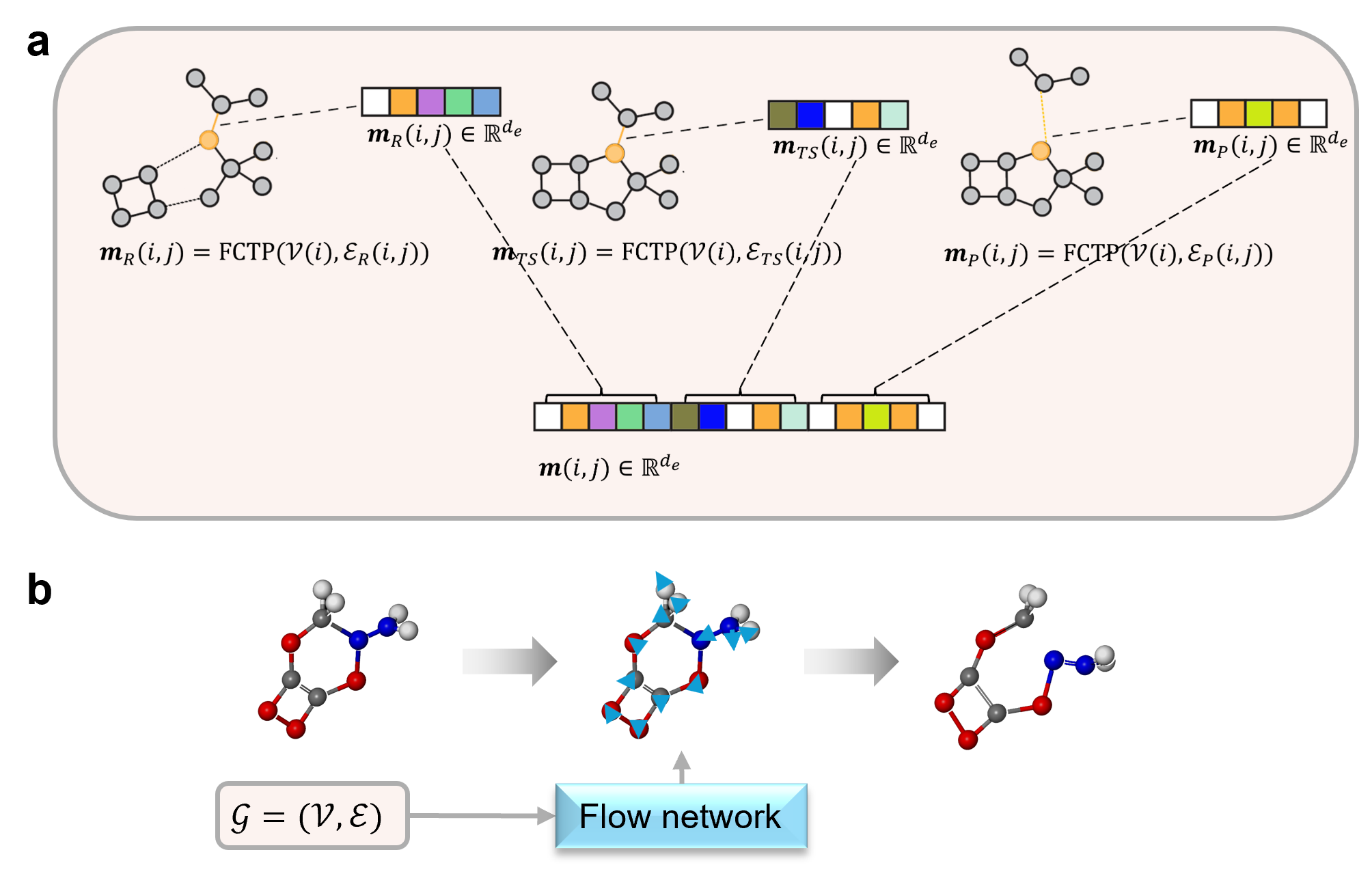}
    \caption{\textbf{Overview of MolGEN.} \textbf{a} Schematic illustration of the reaction message, $\boldsymbol{m}(i,j)$. The molecular messages of the reactants (R) and products (P), denoted as $\boldsymbol{m}_R(i,j)$ and $\boldsymbol{m}_P(i,j)$, are computed from their respective molecular geometries. The transition-state (TS) message, $\boldsymbol{m}_{\rm TS}(i,j)$, is obtained from the flow-matching interpolants. The overall reaction message, $\boldsymbol{m}(i,j)$, is constructed by concatenating $\boldsymbol{m}_R(i,j)$, $\boldsymbol{m}_{\rm TS}(i,j)$, and $\boldsymbol{m}_P(i,j)$.
    \textbf{b} The flow network receives the graph of R, P, and the TS interpolants as inputs, and outputs the velocity field that drives the transformation of the TS interpolant from Gaussian noise to the target geometry. }
    \label{fig:algo}
\end{figure}

MolGEN extends this framework by introducing a conditioning mechanism that steers the generation process toward specific target distributions. MolGEN employs a message passing neural network (MPNN) architecture. For the task of TS generation, we modify the MPNN by concatenating the messages of the transition state (TS) interpolant and the reaction-product pair (RP) to form a reaction message, as illustrated in Figure~\ref{fig:algo}. This process integrates the bond information of all three states into the reaction messages. Analogously, for the task of reaction product generation, we concatenate the messages of the reactant and the product interpolant to form the reaction message.
Subsequently, the model passes these reaction messages through message passing layers to predict the velocity field that evolves the distribution. 

MolGEN initiates with a standard Gaussian distribution. Therefore, it also incorporates stochasticity while performing a deterministic forward pass. The probability path simulated by MolGEN resembles that of a diffusion model, in the sense that both follow a transformation from a standard Gaussian distribution to a jagged target distribution. Consequently, MolGEN is capable of exploring the free energy surface as effectively as diffusion-based models.

We train MolGEN by two different loss functions. The first one is the flow matching loss
\begin{align}\label{eq:L_FM}
    \mathcal{L}_{\rm FM}(\theta)=&\mathbb{E}_{t,p_t(\boldsymbol{x})}\Vert \boldsymbol{v}^{\theta}_t(\boldsymbol{x}) -\boldsymbol{u}_t(\boldsymbol{x})\Vert_1,
\end{align}
where $\theta$ denotes the learnable parameters of the velocity field model and $t\in[0,1]$. $\boldsymbol{x}\sim p_t(\boldsymbol{x})$ denotes the optimal transport path evolving from the initial Gaussian distribution to the target distribution, and $\boldsymbol{u}_t(\boldsymbol{x})$ denotes the corresponding velocity field. Since MolGEN generates a distribution, we also use a Kullback-Leibler divergence loss to measure the difference of the predicted probability path from the optimal transport probability path:
\begin{equation}\label{eq:kl}
    \mathcal{L}_{\rm KL}(\theta)=\mathbb{E}_{t,p_t(\boldsymbol{x})}D_{\rm KL}[p_t(\boldsymbol{x})\Vert p_t^\theta(\boldsymbol{x})].
\end{equation}

By design, MolGEN engrosses the strength of both diffusion and flow matching paradigms. Firstly, it performs sampling via an optimal transport path, thus enabling fast inference that requires only $\sim 0.8\ \rm s$ inference time, orders of magnitudes faster than diffusion model. Secondly, it can generate both TS and reaction products with equal diversity and increased accuracy than diffusion models.
Ultimately, MolGEN demonstrates that flow matching is all that is needed for molecular generation.

\section{Results}
\subsection{Generating transition states}~\label{sec:gen-TS}

We first demonstrate that MolGen is effective in generating transition states.
We trained MolGEN on Transition1x~\cite{schreiner2022transition1x}, a dataset that contains paired reactants, TSs and products calculated from climbing-image nudged elastic band method (NEB)~\cite{henkelman2000climbing} obtained with DFT ($\omega$B97x/6-31G(d))~\cite{chai2008systematic,ditchfield1971self}. Transition1x contains 11,961 reactions with product-reactant pairs sampled from the GDB-7 dataset~\cite{ruddigkeit2012enumeration}. Each reaction consists of up to seven heavy atoms, including C, N, O, and 23 atoms in total. We use the same train-validation-test partitioning as reported with the database~\cite{schreiner2022transition1x}. All datasets are stratified by molecular formula such that no two configurations that come from different data splits are constituted of the same atoms. Test and validation data each consist of 5\% of the total data and are chosen such that configurations contain all heavy atoms (C, N, O).

In the following, we compare the performance of MolGEN with that of diffusion models TSDiff and OA-ReactDiff, as well as a double-ended flow matching model~\cite{duan2025optimal} where a custom prior is designed for each target configuration. MolGEN generates a stochastic distribution by a deterministic pass, indicating two key characteristics. (1) The generation occurs near stationary points on the potential energy surface, where the energy gradient approaches zero. This contrasts with the double-ended flow matching model, which yield a unique configuration. Considering this, we generate 30 independent samples for each test TS and select the one with the lowest energy as the final prediction. Since these samples can be generated in parallel, this approach does not necessarily increase inference time. (2) MolGEN is also capable of identifying alternative TS configurations, including those with lower transition barriers than the reference data, an ability that double-ended flow-matching models lack.

\begin{table}[htbp]
  \centering
  \begin{threeparttable}
  \rowcolors{2}{white}{teal!15}
  \caption{Summary of statistics for RMSD, barrier height error $|\Delta E|$, ratio of valid TS and ratio of updated TS, where the generated TS have lower energies than the corresponding reference TS by more than $0.1\ \rm kcal/mol$.}
    \begin{tabular}{l|l|l|l|l|l|l|l}
    \toprule
    \makecell[l]{Training and \\testing database} & Model & Loss function & \makecell[l]{mean\\ RMSD \\($\rm \mathring{A}$)} & \makecell[l]{mean\\$|\Delta E|$\\$(\rm kcal/mol)$} & \makecell[l]{Chem.\\ accu. \\(\%)} & \makecell[l]{Ratio of \\ valid TS \\ (\%)} & \makecell[l]{Ratio of\\ updated\\ TS (\%)} \\
    \midrule
    \makecell[l]{Transition1x\tnote{a}\\Grambow\tnote{b}} & TSDiff & KL div & 0.253\tnote{a} & 10.369\tnote{a} & 10.9\tnote{a} & 97.00\tnote{b} & 30.96\tnote{b} \\

    Transition1x & \makecell[l]{OA-ReactDiff\\+recommender} & KL div & 0.129\tnote{a} & 4.453\tnote{a} & 48.7\tnote{a} &       &   \\

    Transition1x & React-OT & FM & 0.103\tnote{a} & 3.337\tnote{a} & 63.6\tnote{a} &       &     \\

    \makecell[l]{Transition1x\\(Pretrained \\by RGD1-xTB)} & \makecell[l]{React-OT\\+pretraining} & FM & \makecell[l]{\textbf{0.088}\\0.098\tnote{a}} &  \makecell[l]{\textbf{3.608}\\2.864\tnote{a}} & \makecell[l]{\textbf{57.0}\\71.9\tnote{a}} & \textbf{74.67} & \textbf{4.00}  \\

    Transition1x & MolGEN & \makecell[l]{FM \& \\finetune by KL div} & \textbf{0.106} & \textbf{3.285} & \textbf{54.38} & \textbf{87.46} & \textbf{7.32} \\

    \makecell[l]{RGD1+\\Transition1x} & MolGEN & \makecell[l]{FM \& \\finetune by KL div} & \textbf{0.100} & \textbf{2.759} & \textbf{70.80} & \textbf{89.20} & \textbf{27.87} \\

    \makecell[l]{RGD1+\\Transition1x} & MolGEN & KL div & \textbf{0.112} & \textbf{2.844} & \textbf{68.98} & \textbf{88.50} & \textbf{30.66} \\
    \bottomrule
    \end{tabular}%
    \begin{tablenotes}
    \footnotesize
    \item All tested values in this work are highlighted in bold. React-OT was evaluated using the same pretrained model provided by the original authors. 
    \item [a] The values are borrowed from Duan~\cite{duan2025optimal}, who used Transition1x database~\cite{schreiner2022transition1x} for training and testing.
    \item [b] The values are borrowed from Kim~\cite{kim2024diffusion}, who used the database reported by Grambow \textit{et al.}~\cite{grambow2020reactants} for training and testing.
    \end{tablenotes}
  \label{tab:errors}%
\end{threeparttable}
\end{table}%

MolGEN achieves a mean root mean square deviation (RMSD) of approximately $0.106\ \rm \mathring{A}$ between the generated and reference TS structures on the Transition1x test set, halving the error of diffusion models and matching the error of the double-ended flow matching model~\cite{duan2025optimal} 
(Table~\ref{tab:errors}). The mean barrier height error produced by MolGEN is about $3.285 \rm\ kcal/mol$, also halving that of diffusion models and comparable to that of the double-ended flow matching model ($3.608\ \rm kcal/mol$). 
In chemical reactions, a difference of one order of magnitude in reaction rate is often used as the characterization of chemical accuracy, which translates to $1.58\ \rm kcal/mol$ in barrier height error assuming a reaction temperature of $70\ ^\circ\rm C$. By this standard, over $54\%$ of TS from MolGEN have high chemical accuracy relative to the reference data, on par with that of the state-of-the-art of the double-ended flow matching model ($57\%$).

MolGEN is capable of identifying alternative TS configurations beyond those present in the reference dataset. To quantify this capability, we performed quantum chemical validation on the predicted configurations, using saddle point optimization. The ratio of valid TS, those with a single imaginary vibration frequency, exceeds $87\%$, notably higher than the ratio achieving chemical accuracy. Moreover, over $7\%$ of the TS configurations have energies lower than their corresponding reference TS by more than $0.1\ \rm kcal/mol$.

\paragraph{Training on additional reactions further improves MolGEN.}
The performance of MolGEN can be further enhanced by training on a larger and more diverse dataset. We combined RGD1~\cite{zhao2023comprehensive} with Transition1x for this purpose. RGD1 covers 177,992 reactions involving molecules containing $C$, $H$, $N$, and $O$ atoms, with up to 10 heavy atoms. The TSs and products in RGD1 are derived from NEB obtained with DFT (B3LYP-D3/TZVP). Of these, 3\% and 6\% are held out as a test and validation set, respectively.
Training on this expanded dataset reduced the mean RMSD to $0.100\ \rm \mathring{A}$ and the mean barrier height error to $2.759\ \rm kcal/mol$. More notably, the ratio of valid TS increased to $89.20\%$, while the ratio of TS with energies lower than reference rose sharply to $27.87\%$, approaching the performance achieved by the diffusion-based model TSDiff~\cite{kim2024diffusion} ($30.96\%$). 

\paragraph{Training on KL divergence improves TS identification.} We can drive the model to seek stationary point even further through training only by KL divergence loss, as evidenced in Table~\ref{tab:errors}. When trained only by KL divergence, the ratio of TS with energies lower than reference rose to $30.66\%$ matching the performance of diffusion models~\cite{kim2024diffusion}, even though the mean RMSD increased slightly to $0.112\ \rm \mathring{A}$. This indicates that the model is locating more alternative TS with lower energies than the reference data.

\paragraph{Analysis on failure cases}
We performed a saddle point optimization on the 33 reactions for which the model failed to generate the correct transition state (TS) and verified the validity of the reference TSs by whether they possess a single imaginary frequency. Six of the reference TSs did not pass this saddle point validation.
Among the remaining 26 reactions, 23 failed the intrinsic reaction coordinate (IRC) calculation, indicating the presence of a substantial portion of erroneous data in the database, where the transition states do not match with their corresponding reactions.

\paragraph{Inference time} Since the ODE follows an optimal transport trajectory, it theoretically requires only two evaluations to reach the same target state. To quantify the deviation of the learned dynamics from this optimal transport path, we systematically reduce the number of function evaluations (nfe) during sampling. Empirically, MolGEN attains convergence at nfe=10, at which point the mean RMSD of the generated TS stabilizes, corresponding to an inference time of $0.8\ \rm s$. In practice, we employ the dopri5 adaptive integrator~\cite{hairer1993solving} to guarantee numerical stability and convergence, yielding an average inference time of $2.1\ \rm s$.

\subsection{Generating reaction products}\label{sec:gen-p}
In the TS generation task, MolGEN was used to realize a one-to-one mapping, i.e., from the reactant-product pair to the TS. Here, we demonstrate that MolGEN can also capture open-ended mappings with multiple outcomes, by employing MolGEN to generate the product from given reactant.

MolGEN was trained on a combined dataset of Transition1x and RGD1, comprising elementary reaction data as described in the previous section. Model performance was evaluated on the Transition1x test set.

We assessed predictive performance using top-k accuracy, which quantifies the model’s ability to correctly anticipate plausible products for a given reactant. Leveraging the distributional nature of conditional flow matching, MolGEN can be sampled repeatedly for the same reactant to generate a diverse ensemble of candidate products. Each predicted product geometry is converted to an adjacency matrix~\cite{seo2021topology} and ranked according to its sampling frequency.

The top-k step accuracy measures the fraction of individual elementary steps where the recorded product appears within the top-k ranked predictions. Multiple valid products can emerge from the same reactants due to differences in reaction conditions (e.g., the use of distinct acids, bases, or catalysts) or alternative resonance structures. Within the Transition1x test set, we identified 37 unique reactants (distinct adjacency matrix), each associated with between one and twenty-seven distinct products, totaling 287 reaction entries.

We compare MolGEN’s top-k accuracy against that of a prototypical sequence-based model, Graph2SMILES~\cite{tu2022permutation}, and a sequence model with electron distribution encoding, FlowER~\cite{joung2025electron}. While MolGEN achieves a lower top-1 accuracy (43\%) compared to Graph2SMILES (80\%), its top-3 (and larger k) accuracy rises sharply to near 100\% (Figure~\ref{fig:rn-khp}a), on par with FlowER and surpassing Graph2SMILES, which saturates at 90\%. Notably, since MolGEN preserves atomic identities and stoichiometry, it inherently avoids the hallucinatory artifacts often observed in sequence-based models, where mass or electron conservation can be violated. 

To further demonstrate of MolGEN's ability, in the next section, we combine the product generation model and the TS generation model obtained in the previous section to address the task of reaction network generation.

\subsection{Toward a Self-Contained Solution for Reaction Network Generation}\label{sec:rng}
By integrating the product generation model with the transition-state (TS) generation model, we establish a self-contained framework for automated reaction network generation.
A key motivation for developing predictive chemistry models that operate at the mechanistic (elementary step) level is their potential to generalize beyond standard reaction types encountered during training. To evaluate this capability, we apply our method to the decomposition network of $\gamma$-ketohydroperoxide (KHP), a realistic and chemically significant system whose reactant does not appear in the training dataset. 

\begin{figure}
    \centering
    \includegraphics[width=1\linewidth]{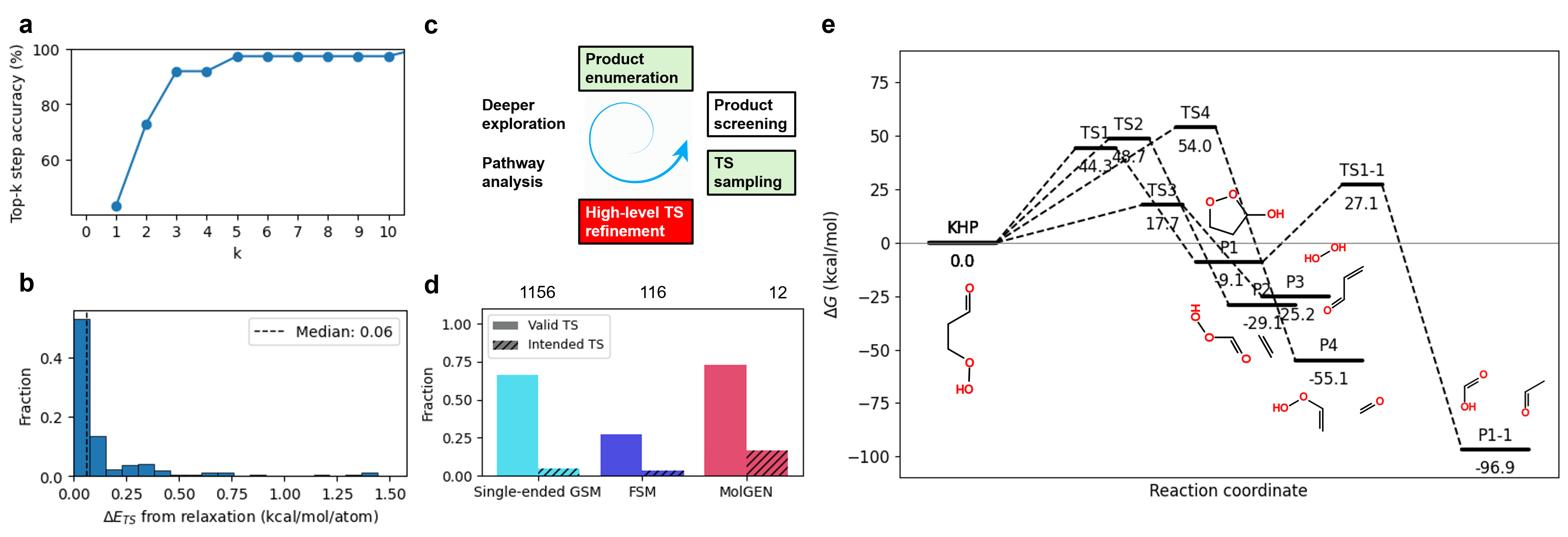}
    \caption{\textbf{a} Top-$k$ step accuracy for product prediction in a single elementary reaction. \textbf{b} Energy change of the transition state ($\Delta E_\mathrm{TS}$) following structural relaxation. \textbf{c} Workflow of reaction network exploration. \textbf{d} Fraction of valid and intended transition states (TS) obtained using three methods: MolGEN, single-ended growing string method (GSM), and freezing string method (FSM). The average number of gradient descents per TS optimization is indicated above each bar. Values for single-ended GSM and FSM are adapted from Grambow~\cite{grambow2018unimolecular}. \textbf{e} Representative degradation pathways of KHP. All energies are computed at the $\omega$B97X/6-31G(d) level of theory.}
    \label{fig:rn-khp}
\end{figure}

KHP decomposition proceeds through multiple mechanistic steps. Figure~\ref{fig:rn-khp}c illustrates a representative workflow for reaction network exploration, comprising five stages: (1) product enumeration, (2) product pre-pruning, (3) TS sampling, (4) high-level TS refinement including IRC calculations to validate that TS geometries correspond to the given reaction, and (5) reaction relevance analysis. These stages are recursively applied to both the initial reactant(s) and any intermediates generated during exploration. The computational bottleneck lies in the TS refinement stage; consequently, extensive efforts focus on optimizing product and TS sampling and pruning to minimize the number of expensive TS refinements required.

For product generation, traditional reaction network exploration packages such as Yet Another Reaction Program (YARP) \cite{zhao2021simultaneously,zhao2022algorithmic} employ the graph enumeration to explore potential products based on empirical rules like b2f2 (break 2 bonds and form 2 new bonds). While effective, such rules may overlook feasible products.
TS sampling presents an additional challenge. Since reactant-product pairs can be aligned in infinitely many ways, the alignment influences the success rate of subsequent TS refinements. YARP employs heuristic criteria, such as mass-weighted root-mean-square displacement, to identify alignments likely to yield successful TS optimizations via nudged elastic band (NEB) or growing string method (GSM). 

In our framework, the product generation model performs stage (1), while the TS generation model handles stage (3). Using KHP as input, the product generation model generated 1000 candidate products, which were grouped into 80 unique products based on adjacency matrices. Geometry optimizations of the lowest-energy configurations identified 65 local minima, as verified by vibrational analysis, six of which shared the same adjacency matrix as the original KHP structure. The remaining 59 unique products were used as inputs for the TS generation model, which generated 30 TS candidates per product. The lowest-energy candidate for each product was selected as the TS initial guess.
Out of the 59 selected TS guesses, 40 were successfully optimized to valid TS structures characterized by a single imaginary frequency. These TS were subjected to IRC calculations, yielding 14 intended reactions with intended transition states. Among the products of these intended reactions, three contained radicals and were excluded, leaving the remaining 11 products as reactants for subsequent exploration steps.

The fractions of valid and intended transition-state (TS) structures, along with the average number of gradient descent calls (excluding IRC calculations) required for TS optimization, are summarized in Figure~\ref{fig:rn-khp}d and compared with results from string-based methods reported by Grambow \textit{et al.}~\cite{grambow2018unimolecular}. Both the valid and intended TS fractions achieved by MolGEN exceed those of the string methods. Moreover, the number of gradient descent calls is drastically reduced to 12 per reaction, indicating that the TS geometries generated by MolGEN lie in close proximity to the corresponding TS. The associated energy changes following TS relaxation are shown in Figure~\ref{fig:rn-khp}b.

\section{Discussion and conclusion}
Flow matching can establish a mapping from a simple prior distribution to a complex data distribution by an optimal-transport path. We develop a conditional flow matching framework, MolGEN, capable of addressing both double-ended generation tasks (one-to-one mappings), such as transition-state (TS) generation from known reactant-product pairs, and open-ended generation tasks (multiple-target sampling), such as product generation from a known reactant. In contrast to diffusion-based models, which rely on computationally expensive stochastic sampling, flow matching enables sub-second inference, offering a significant speed advantage without compromising accuracy and diversity.

Unlike sequence-based models, the MolGEN product generator does not suffer from hallucination artifacts and achieves state-of-the-art top-$k$ accuracy. Furthermore, it provides high-quality initial geometries for products, exhibiting a median post-relaxation energy change of only $0.19\ \mathrm{kcal/mol}$ per atom, indicating strong alignment between predicted and equilibrium structures.

By integrating the TS generating model and product generating model, MolGEN offers a self-contained framework for automated reaction network generation. The fraction of both valid and intended TSs is substantially improved relative to traditional methods, while the optimization cost is drastically reduced. Nonetheless, not all intended TSs corresponding to the predicted products were recovered in our demonstration. This implies two key points: 1) The model requires further refinement, particularly through the use of a higher-quality database, as the current one contains a high proportion of unintended TS; 2) Complementary strategies, such as extensive reactant-product alignment sampling before input into the TS-generating model, or the exclusion of small molecules from reactants to steer exploration toward decomposition pathways, remain valuable for exhaustive searches.

In summary, MolGEN illustrates that flow matching provides a unified, accurate, and computationally efficient foundation for both double-ended and open-ended molecular generation. Its combination of enhanced precision, deterministic inference, and scalability points toward a promising future for flow-matching models as the method of choice for molecular and reaction generation.

\section{Methods}\label{sec:methods}
\subsection{Flow matching by optimal transport path}
\subsubsection{Preliminaries: continuous normalizing flows}

Let $\mathbb{R}^d$ denote the data space with points $\boldsymbol{x}\in\mathbb{R}^d$. Two important variables we use are: the probability density path $p_t\in [0,1]$, which is a time-dependent density, and a time-dependent vector field $\boldsymbol{v}_t\in\mathbb{R}^d$. The vector field $\boldsymbol{v}_t$ generates a time-dependent diffeomorphism (or a flow) $\phi_t:\mathbb{R}^d\rightarrow\mathbb{R}^d$ via the ordinary differential equation (ODE):
\begin{align}
\frac{d}{dt}\phi_t(\boldsymbol{x})&=\boldsymbol{v}_t\bigl(\phi_t(\boldsymbol{x})\bigr),\label{eq:ode1}\\
\phi_0(\boldsymbol{x})&\sim p_0.\label{eq:ode2}
\end{align}

Chen \textit{et al.}~\cite{chen2018neural} suggested modeling $\boldsymbol{v}_t$ by a neural network $\boldsymbol{v}_t^{\theta}(\boldsymbol{x})$, where $\theta$ are learnable parameters, yielding a deep parametrization of the flow $\phi_t$, called a continuous Normalizing Flow (CNF). A CNF pushes a simple base density $p_0$ (e.g.\ Gaussian noise) to a complex target $p_1$ via
\begin{equation}
    p_t =[\phi_t]_* p_0,
\end{equation}
where the push-forward operator is defined by the change-of-variables formula
\begin{equation}
    [\phi_t]_* p_0(\boldsymbol{x})
    = p_0\bigl(\phi_t^{-1}(\boldsymbol{x})\bigr)\det\!\Bigl(\partial_{\boldsymbol{x}}\phi_t^{-1}(\boldsymbol{x})\Bigr).
\end{equation}

A vector field $\boldsymbol{v}_t$ generates a density path $p_t$ if its flow $\phi_t$ satisfies the above. Equivalently, one may verify the continuity equation
\begin{equation}
    \partial_t p_t + \nabla\!\cdot\bigl(p_t\boldsymbol{v}_t\bigr)=0.
\end{equation}

\subsubsection{Flow matching}

Let $\boldsymbol{x}_1$ be a random variable drawn from an unknown data distribution $q(\boldsymbol{x}_1)$. We assume that while samples from $q(\boldsymbol{x}_1)$ are available, the density function itself is inaccessible. Let $p_0$ denote a simple base distribution, such as the standard normal $p_0(\boldsymbol{x}) = \mathcal{N}(\boldsymbol{x}|0, \boldsymbol{I})$, and let $p_1$ be a distribution that closely approximates $q$. The flow matching objective is then formulated to learn a continuous transformation or probability path, that transports samples smoothly from $p_0$ to $p_1$.

A simple way to construct a target probability path is via a mixture of simpler conditional paths~\cite{lipman2022flow}. Given a data sample $\boldsymbol{x}_1$, we denote by $p_t(\boldsymbol{x}|\boldsymbol{x}_1)$ a conditional probability path such that it satisfies $p_0(\boldsymbol{x}|\boldsymbol{x}_1)=p_0$ at $t=0$, and we design $p_1(\boldsymbol{x}|\boldsymbol{x}_1)$ to be concentrated around $\boldsymbol{x}_1$ (e.g., $p_1(\boldsymbol{x}|\boldsymbol{x}_1)=\mathcal{N}(\boldsymbol{x}|\boldsymbol{x}_1,\sigma^2\boldsymbol{I})$ for small $\sigma>0$). Marginalizing over the unknown data distribution $q(\boldsymbol{x}_1)$ gives the marginal probability path~\cite{lipman2022flow}
\begin{equation}
    p_t(\boldsymbol{x})=\int p_t(\boldsymbol{x}|\boldsymbol{x}_1)q(\boldsymbol{x}_1)\mathrm{d}\boldsymbol{x}_1,
\end{equation}
which at $t=1$ yields
\begin{equation}
    p_1(\boldsymbol{x})=\int p_1(\boldsymbol{x}|\boldsymbol{x}_1)q(\boldsymbol{x}_1)\mathrm{d}\boldsymbol{x}_1\approx q(\boldsymbol{x}).
\end{equation}

One can also define a marginal vector field by “mixing” the conditional fields~\cite{lipman2022flow}:
\begin{equation}
    \boldsymbol{v}_t(\boldsymbol{x})
    =\int \boldsymbol{u}_t(\boldsymbol{x}|\boldsymbol{x}_1)\frac{p_t(\boldsymbol{x}|\boldsymbol{x}_1)q(\boldsymbol{x}_1)}{p_t(\boldsymbol{x})}\mathrm{d}\boldsymbol{x}_1,
\end{equation}
where $\boldsymbol{u}_t(\boldsymbol{x}|\boldsymbol{x}_1)$ generates the conditional path $p_t(\boldsymbol{x}|\boldsymbol{x}_1)$. This aggregation yields the correct vector field for the marginal path $p_t(\boldsymbol{x})$.

\subsubsection{Gaussian probability path}

The flow matching objective works with any choice of probability path and vector field. We can construct $p_t(\boldsymbol{x} |\boldsymbol{x}_1)$ and $\boldsymbol{u}_t(\boldsymbol{x} |\boldsymbol{x}_1)$ for a general family of Gaussian conditional paths~\cite{lipman2022flow}. Namely, we set
\begin{equation}\label{eq:cond-path}
p_t(\boldsymbol{x} |\boldsymbol{x}_1)
= \mathcal{N}\bigl(\boldsymbol{x} |\boldsymbol{\mu}_t(\boldsymbol{x}_1),\sigma_t(\boldsymbol{x}_1)^2 \boldsymbol{I}\bigr),
\end{equation}
where $\boldsymbol{\mu}_t\in\mathbb{R}^d$ is a time-dependent mean and $\sigma_t\in(0,1]$ a time-dependent standard deviation. We require $\boldsymbol{\mu}_0(\boldsymbol{x}_1)=0$, $\sigma_0(\boldsymbol{x}_1)=1$ so that $p_0(\boldsymbol{x}|\boldsymbol{x}_1)=\mathcal{N}(\boldsymbol{x}|0,\boldsymbol{I})$ and $\boldsymbol{\mu}_1(\boldsymbol{x}_1)=\boldsymbol{x}_1$, $\sigma_1(\boldsymbol{x}_1)=\sigma_{\min}$ with $\sigma_{\min}\ll1$. Thus, $p_1(\boldsymbol{x}|\boldsymbol{x}_1)$ concentrates at $\boldsymbol{x}_1$.

Among the infinitely many vector fields generating this path, Lipman \textit{et al.} proposed to choose the simplest affine flow conditioned on $\boldsymbol{x}_1$~\cite{lipman2022flow}:
\begin{equation}\label{eq:psi-flow}
    \psi_t(\boldsymbol{x})=\sigma_t(\boldsymbol{x}_1)\boldsymbol{x} +\boldsymbol{\mu}_t(\boldsymbol{x}_1),
\end{equation}
which pushes the standard normal $p_0(\boldsymbol{x}|\boldsymbol{x}_1)=\mathcal{N}(\boldsymbol{x}|0,\boldsymbol{I})$ to $p_t(\boldsymbol{x}|\boldsymbol{x}_1)$ via
\begin{equation}\label{eq:push-forward}
    [\psi_t]_*p_0(\boldsymbol{x})=p_t(\boldsymbol{x}|\boldsymbol{x}_1).
\end{equation}
The associated conditional vector field then follows from differentiating the flow:
\begin{equation}\label{eq:ut-conditional}
    \frac{d}{dt}\psi_t(\boldsymbol{x})=\boldsymbol{u}_t\bigl(\psi_t(\boldsymbol{x})|\boldsymbol{x}_1\bigr).
\end{equation}
Reparameterizing $p_t(\boldsymbol{x}|\boldsymbol{x}_1)$ in terms of just $\boldsymbol{x}_0$ gives the conditional flow matching (CFM) loss:
\begin{equation}\label{eq:cfm-loss}
    \mathcal{L}_{\mathrm{CFM}}(\theta)=\mathbb{E}_{t\sim\mathcal{U}[0,1],q(\boldsymbol{x}_1),p_0(\boldsymbol{x}_0)}\Bigl\|\boldsymbol{v}_t^{\theta}\bigl(\psi_t(\boldsymbol{x}_0)\bigr)-\boldsymbol{u}_t(\psi_t(\boldsymbol{x}_0)|\boldsymbol{x}_1)\Bigr\|^2,
\end{equation}
where $\theta$ are learnable parameters of a neural network model. $\mathcal{U}[0,1]$ is a uniform distribution over the interval $[0,1]$. The CFM loss have identical gradients w.r.t. $\theta$ as the FM loss Eq.~\ref{eq:L_FM}, but is simpler to compute. Thus, it is the common choice for flow matching training~\cite{lipman2022flow}.

In practice, we use a regression based objective~\cite{albergo2023stochastic}
\begin{equation}
    \mathcal{L}_\mathrm{CFM}^{\prime}[\theta]=\mathbb{E}_{t\sim\mathcal{U}[0,1],q(\boldsymbol{x}_1),p_0(\boldsymbol{x}_0)}\Big[\frac{1}{2}|\boldsymbol{v}^{\theta}_t(\psi_t(\boldsymbol{x}_0))|^2-\boldsymbol{u}_t(\psi_t(\boldsymbol{x}_0)|\boldsymbol{x}_1)\cdot\boldsymbol{v}_t^{\theta}(\psi_t(\boldsymbol{x}_0)) \Big],
    \label{eq:loss-b}
\end{equation}
which yields more stable training.

Given the Gaussian probability path $p_t(\boldsymbol{x}| \boldsymbol{x}_1)$ in equation \eqref{eq:cond-path}, and the corresponding flow map $\psi_t$ in equation \eqref{eq:psi-flow}, the conditional vector field Eq.~\ref{eq:ut-conditional} has the form
\begin{equation}\label{eq:ut-gaussian}
    \boldsymbol{u}_t(\boldsymbol{x}| \boldsymbol{x}_1)=\frac{\sigma_t'(\boldsymbol{x}_1)}{\sigma_t(\boldsymbol{x}_1)}\bigl(\boldsymbol{x} - \boldsymbol{\mu}_t(\boldsymbol{x}_1)\bigr)+\boldsymbol{\mu}_t'(\boldsymbol{x}_1).
\end{equation}

\subsubsection{Optimal transport path}
A natural choice for probability paths is to define the mean and the std to simply change linearly in time:
\begin{equation}\label{eq:mu-sigma}
    \boldsymbol{\mu}_t(\boldsymbol{x}_1) = t\boldsymbol{x}_1,\qquad\sigma_t(\boldsymbol{x}_1) = 1 - \bigl(1 - \sigma_{\min}\bigr)t.
\end{equation}
Then this path is generated by the velocity field~\cite{lipman2022flow}
\begin{equation}\label{eq:ut-specific}
    \boldsymbol{u}_t(\boldsymbol{x} | \boldsymbol{x}_1)= \frac{\boldsymbol{x}_1 - \bigl(1 - \sigma_{\min}\bigr)\boldsymbol{x}}{1 - \bigl(1 - \sigma_{\min}\bigr)t}.
\end{equation}
The conditional flow that corresponds to $\boldsymbol{u}_t(\boldsymbol{x}|\boldsymbol{x}_1)$ is
\begin{equation}
    \psi_t(\boldsymbol{x})=t\boldsymbol{x}_1+(1-(1-\sigma_{\rm min})t)\boldsymbol{x}.
\end{equation}
Allowing the mean and std to change linearly not only leads to simple and intuitive paths, but it is actually also optimal in the following sense. The conditional flow $p_t(\boldsymbol{x}|\boldsymbol{x}_1)$ is in fact the optimal transport displacement map between the two Gaussians $p_0(\boldsymbol{x}|\boldsymbol{x}_1)=\mathcal{N}(\boldsymbol{x}|0,\boldsymbol{I})$ and $p_1(\boldsymbol{x}|\boldsymbol{x}_1)$.

Using the conditional probability Eq.~\ref{eq:cond-path}, we can evaluate the probability of a model predicted path by
\begin{equation}\label{eq:pred-conditional-probpaht}
    p^{\theta}_t(\boldsymbol{x}|\boldsymbol{x}_1)=\frac{1}{\sqrt{2\pi\sigma_t^2}}\text{exp}\left[-\frac{(\boldsymbol{x} - \boldsymbol{\mu}^\theta_t)^2}{2\sigma_t^2} \right],
\end{equation}
where the expectation of $\boldsymbol{\mu}^\theta_t$ can be evaluated by substituting the model predicted velocity field $\boldsymbol{v}^\theta_t(\psi_t(\boldsymbol{x}_0))$ in Eq.~\ref{eq:ut-specific}:
\begin{align}\label{eq:pred-x}
    \nonumber \boldsymbol{\mu}^\theta_t =& \frac{ \boldsymbol{x}_1-(1-(1-\sigma_{\rm min})t)\boldsymbol{v}^\theta_t(\psi_t(\boldsymbol{x}_0))}{1-\sigma_{\rm min}}\\
    =& t\boldsymbol{v}^\theta_t(\psi_t(\boldsymbol{x}_0)).
\end{align}
Substituting Eq.~\ref{eq:mu-sigma} and Eq.~\ref{eq:pred-x} in Eq.~\ref{eq:pred-conditional-probpaht} yields the probability of a model predicted path:
\begin{align}\label{eq:pred-condprobpath-specific}
    p^{\theta}_t(\boldsymbol{x}|\boldsymbol{x}_1)= \frac{1}{\sqrt{2\pi(1-(1-\sigma_{\rm min})t)^2}}\text{exp}\left[ -\frac{\left[\boldsymbol{x}-t\boldsymbol{v}^\theta_t(\psi_t(\boldsymbol{x}_0))\right]^2}{2(1-(1-\sigma_{\rm min})t)^2} \right]
\end{align}

\subsubsection{KL divergence loss}
Besides the CFM loss, we also use a KL divergence loss to measure the the discrepancy between the predicted probability path and the optimal transport probability path. Similar to the FM loss, the KL divergence loss Eq.~\ref{eq:kl} is intractable. Instead, we propose a conditional KL divergence loss:
\begin{align}\label{eq:kl-expand}
    \mathcal{L}_{CKL}(\theta,\boldsymbol{x})=&\mathbb{E}_{t,q(\boldsymbol{x}_1),p_0(\boldsymbol{x}_0)}D_{\rm KL}(p_t(\boldsymbol{x}| \boldsymbol{x}_1)\Vert p^\theta_t(\boldsymbol{x}|\boldsymbol{x}_1)) \nonumber \\
    =&\mathbb{E}_{t,q(\boldsymbol{x}_1),p_0(\boldsymbol{x}_0)}\frac{1}{\sqrt{2\pi(1-(1-\sigma_{\rm min})t)^2}} \text{exp}\left( -\frac{\left[\boldsymbol{x}-\boldsymbol{\mu}_t(\boldsymbol{x}_1)\right]^2}{2(1-(1-\sigma_{\rm min})t)^2} \right) \nonumber \\
    & \left(-\frac{1}{2(1-(1-\sigma_{\rm min})t)^2} 
    \left( [\boldsymbol{x}-\boldsymbol{\mu}_t(\boldsymbol{x}_1)]^2-[\boldsymbol{x}-t\boldsymbol{v}^\theta_t(\psi_t(\boldsymbol{x}_0))]^2 \right) \right).
\end{align}
In practice, we set the random position $\boldsymbol{x}$ to $\boldsymbol{0}$. To mitigate the numerical instability caused by the denominator term $(1-(1-\sigma_{\rm min})t)$, we make two modifications to Eq.~\ref{eq:kl-expand}. Firstly, we drop the normalizing factor $\frac{1}{\sqrt{2\pi(1-(1-\sigma_{\rm min})t)^2}}$, which only affects the relative weighting of the loss across different $t$. Secondly, we compute the KL divergence between two alternative Gaussian probabilities:
\begin{equation}
    \mathcal{L}_{CKL}^\prime (\theta) = \mathbb{E}_{t,q(\boldsymbol{x}_1),p_0(\boldsymbol{x}_0)}\text{exp}\left( -\frac{[t\boldsymbol{\mu}_t(\boldsymbol{x}_1)]^2}{2}\right)\left( -\frac{1}{2}\left([t\boldsymbol{\mu}_t(\boldsymbol{x}_1)]^2 - [t\boldsymbol{v}^\theta_t(\psi_t(\boldsymbol{x}_0))]^2 \right)\right),
\end{equation}
where both distributions share the same mean as the actual probability path but have a standard deviation of $1/t$, which remains strictly positive.

While KL divergence is a measure of how different two distributions are, it is not a metric. In particular, it is not symmetric in the two distributions and does not satisfy the triangle inequality. Therefore, we use a symmetric variant, the Jensen Shannon divergence. Let $P$ be the reference distribution, and $Q$ be the predicted distribution, Jensen Shannon divergence is defined by
\begin{equation}
    D_{JS}=\frac{1}{2}D_{\rm KL}(P\Vert M)+\frac{1}{2}D_{\rm KL}(Q\Vert M),\ M=\frac{1}{2}(P+Q).
\end{equation}
The Jensen Shannon divergence is a metric.

\subsubsection{Dynamic optimal transport}
Dynamic optimal transport, as introduced by Benamou and Brenier~\cite{benamou2000computational}, presents the squared Wasserstein-2 distance $W^2_2(p_0,p_1)$ in a dynamic formulation. This formulation involves optimizing the squared $L^2$ norm of a time-varying vector field $\boldsymbol{u}_t(\boldsymbol{x})$, which transports samples from $\boldsymbol{x}_0 \sim p_0(\boldsymbol{x}_0)$ to $\boldsymbol{x}_1 \sim p_1(\boldsymbol{x}_1)$ according to an ordinary differential equation $\frac{d\psi_t(\boldsymbol{x})}{dt}=u_t(\boldsymbol{x})$. Mathematically, this dynamic formulation is expressed as
\begin{equation}
    W^2_2(p_0,p_1):=\inf_{\pi\in\Pi(p_0,p_1)}\int_0^1\int_{\mathbb{R}^d} \frac{1}{2}\Vert \boldsymbol{u}_t(x) \Vert^2p_t(\boldsymbol{x})\text{d}\boldsymbol{x}\text{d}t,
\end{equation}
where $\Pi (p_0,p_1)$ is the set of couplings with $p_0$ and $p_1$.
This approach is in contrast with the flow matching method by Lipman et al.~\cite{lipman2022flow}, where the probability path is given by Gaussian distributions.
In ReactOT~\cite{duan2025optimal}, Duan et al. assumed the optimal coupling to be $\pi^*\sim(\frac{R+P}{2},TS)$. This enables a flow matching model that generates a unique TS for a given reactant-product pair.

\subsection{MPNN}\label{sec:method-gnn}

An atomic graph is embedded in 3-dimensional (3D) Euclidean space, where each node represents an atom and edges connect nodes if the corresponding atoms are within a given cutoff distance of each other. The cutoff was set to $12\ \rm \mathring{A}$. We represent the state of each node $i$ by a tuple $\sigma_i=(\boldsymbol{r}_i,z_i,\boldsymbol{h}_i)$, where $\boldsymbol{r}\in\mathbb{R}^3$ is the position of atom $i$, $z_i$ the chemical element, and $\boldsymbol{h}_i$ are learnable features.
MPNN parameterises a mapping from a graph to a target space, such as the velocity field, through multiple message passing layers.

\subsubsection{Node feature}
Firstly, the one-hot coded chemical element $z_i$ is passed through a SiLU-activated multi-layer perception (MLP) to obtain the node feature $\mathcal{V}_i$.

\subsubsection{Construction of edge features}

An edge feature $\mathcal{E}=\oplus^{l_{\rm max}}_{l=0}d_ll^{p(l)}$ is constructed as $d_l$ copies of degree-$l$ spherical components with parity $p(l)= \text{even}$ for even $l$ and odd for odd $l$, so that the edge features transform as proper SO(3) tensors. This mirrors the standard design in higher-order equivariant MPNNs where messages are built by contracting node features with spherical harmonics of edge directions and radial envelopes, then mapped via Clebsch–Gordan (tensor-product) rules to the desired output irreps~\cite{batzner20223}.

\paragraph{Radial basis}
For each edge length $r_{ji}=\Vert \boldsymbol{r}_{ji}\Vert$, we compute a set of radial channels
\begin{equation}
    \phi_k=\frac{1}{2}\left( \cos(\pi r/r_{\rm c})+1 \right)1[r<r_{\rm c}]\cdot \text{exp}\left(-\beta_k(e^{-r}-\nu_k)^2\right),\ k=1,...,K,
\end{equation}
with evenly spaced means $\nu_k$ between $e^{-r_{\rm c}}$ and $1[r<r_{\rm c}]=\begin{cases}
    1, & \text{if } r<r_{\rm c} \\
    0, & \text{if } r>r_{\rm c}
\end{cases}$, and a shared width $\beta_k= \frac{K}{\left(\frac{2}{K}(1-e^{-r_{\rm c}})\right)^2} $. This yields smooth bounded radial features.

\paragraph{Angular basis}
For the unit direction $\hat{\boldsymbol{r}}_{ji}=\boldsymbol{r}_{ji}/r_{ji}$, the block computes real spherical harmonics $Y_l(\hat{\boldsymbol{r}})\in\mathbb{R}^{2l+1}$ up to $l_{\rm max}$. Spherical harmonics are the canonical basis that ensures proper SO(3) transformation.

\paragraph{Per-$l$ mixing and the edge feature tensor}
For each $l$, the radial channels $\phi(r_{ji})\in\mathbb{R}^K$ is mapped via a small SiLU-activated MLP to $d_l$ scalar weights $\textbf{w}_l(r_{ji})\in\mathbb{R}^{d_l}$. These weights scale each $Y_l(\hat{\boldsymbol{r}}_{ji})$, producing 
\begin{equation}
    \textbf{e}_{l}(\boldsymbol{r}_{ji})=\textbf{w}_l \otimes Y_l(\hat{\boldsymbol{r}}_{ji})\in\mathbb{R}^{d_l(2l+1)}.
\end{equation}
Concatenating over $l=0,...,l_{\rm max}$ yields a flat edge feature tensor $\textbf{e}_{ji}$ with irreps $\mathcal{E}_{ji}=\oplus_l d_l l^{p(l)}$.

\subsubsection{Construction of messages by Clebsch-Gordan contraction}
Messages are computed by a learned fully connected tensor product (FCTP) that contracts node features and edge features: $\boldsymbol{\mathcal{M}}(j,i)=\text{FCTP}(\mathcal{V}_i,\textbf{e}_{ji})$,
where $\text{FCTP}$ enumerates all Clebsch-Gordan (CG) paths $\mathcal{V}_i\otimes \mathcal{E}_{ji}$ and learns weights for them while preserving equivariance by construction. The invariant components of the resulting messages, denoted $\boldsymbol{m}(j,i)$, are subsequently extracted and propagated during message passing. 

\paragraph{Reaction messages}
The messages of the reactant, TS, and product configurations are concatenated to form a reaction message:
\begin{equation}
    \boldsymbol{m}(j,i)=\oplus_{R,TS,P}(\boldsymbol{m}_R(j,i), \boldsymbol{m}_{\rm TS}(j,i), \boldsymbol{m}_P(j,i) ).
\end{equation}
Note that in practice, the $\boldsymbol{m}_{\rm TS}(j,i)$ is built using a flow matching sample at time $t$, while $\boldsymbol{m}_{R}(j,i)$ and $\boldsymbol{m}_{P}(j,i)$ are built using the true configurations.

\subsubsection{Object-aware Message passing}
Message passing is implemented with object-aware equivariance, where the model prediction is equivariant w.r.t. the relative rotation and translation of various molecules in a system~\cite{duan2023accurate}. 

For atoms belonging to the same molecule, the concatenated vector of the message and the associated node features, $\oplus(\mathcal{V}_i, \mathcal{V}_j, \boldsymbol{m}(j,i))$, is passed through an SiLU-activated MLP to produce $M_{ji}$. These outputs are then aggregated to form two components: a vector feature, $\boldsymbol{H}_i = \sum_{j \in \mathcal{N}(i)} M_{ji} \boldsymbol{r}_{ji}$, and a scalar feature, $\sum_{j \in \mathcal{N}(i)} M_{ji} \mathcal{V}_j$. The scalar feature is concatenated with $\mathcal{V}_i$ and passed through another SiLU-activated MLP to yield the updated scalar feature $H_i^{intra}$.

For atoms belonging to different molecules, only the node features are aggregated, i.e., $\oplus(\mathcal{V}_i, \mathcal{V}_j)$. This concatenated representation is passed through a SiLU-activated MLP to produce $M^{\text{inter}}_{ji}$. The resulting messages are aggregated as $\sum_{j \in \mathcal{N}(i)} M^{\text{inter}}_{ji} \mathcal{V}_j$, which is then concatenated with $\mathcal{V}_i$ and processed by another SiLU-activated MLP to obtain the inter-molecular scalar feature $H^{\text{inter}}_i$.

The intra- and inter-molecular scalar features are concatenated $\oplus(H_i^{intra}, H^{\rm inter}_i)$ and passed through a SiLU-activated MLP to produce the final updated scalar feature $H_i$.

\subsubsection{Readout by an equivariant transformer}
To efficiently capture the intricate interactions between atoms both within and across molecules, we employ the equivariant transformer (ET)~\cite{jiao2024equivariant} to update the scalar and vector features obtained in the previous section. The ET architecture preserves object-aware SE(3) symmetry, ensuring consistent treatment of rotations and translations, while effectively modeling complex geometric relationships. It exhibits strong scalability across large and diverse datasets, showing substantial performance gains when trained on extended molecular databases, as demonstrated by our results in the main text. This advantage is further reinforced by its computational efficiency, where the ET trains approximately 3–4× faster per epoch than existing architectures such as React-OT.

The ET outputs a vector feature $\boldsymbol{H}^{\prime}_i$ and a scalar feature $H^{\prime}_i$.
Finally, the vector feature $\boldsymbol{H}^{\prime}_i$ is passed through a simple linear layer to produce the predicted velocity field $\boldsymbol{v}_i$.

\subsection{Model training}
We train the model using the AdamW optimizer~\cite{loshchilov2017decoupled}. 
The learning rate is initialized at $1\times10^{-4}$ and linearly warmed up to $5\times10^{-4}$ over the first 10 epochs according to
\begin{equation}
    lr = 1\times10^{-4} + (5\times10^{-4} - 1\times10^{-4}) \times \frac{\text{epoch} + 1}{10},
\end{equation}
and then decays to $3\times10^{-5}$ following a cosine schedule:
\begin{equation}
    lr = 3\times10^{-5} + (5\times10^{-4} - 3\times10^{-5}) \times 0.5 \times \left(1 + \cos\left(\pi \frac{\text{epoch} - 10}{600}\right)\right).
\end{equation}
During finetuning, the learning rate starts at $1\times10^{-4}$ and decays to $3\times10^{-5}$ with the same cosine decay strategy. 
The batch size is set to $64$.

\subsection{Evaluation}
\paragraph{DFT:}
Energy evaluation was performed using DFT ($\omega$B97x/6-31G(d)), for consistent comparison with the Transition1x database. For saddle point optimization, $\omega$B97x/def2-TZVP was used for more reliable convergence.
Energy evaluation was carried out using the pySCF package~\cite{sun2018pyscf}, and optimization was carried out using the Psi4 package~\cite{smith2020psi4}.

\paragraph{IRC calculation:}
IRC performed to verify the optimized TS is carried out using the Psi4 package~\cite{smith2020psi4}. The resulting endpoints of the IRC integration are further optimized to energy minimums. The comparison of IRC endpoints and input reactants and products is based on the adjacency matrix computed by the TAFFI package~\cite{seo2021topology}.

\paragraph{Measurement of top-k accuracy:}
For each reaction entry in the Transition1x test set, we extract the reactant and generate 240 corresponding product candidates. All generated samples are then combined, and unique reactants are identified. For each unique reactant, we compile the list of associated products and rank them based on their occurrence frequency. Molecular comparisons are performed using adjacency matrices computed with the TAFFI package~\cite{seo2021topology}.

\section*{Code availability}
The code base for MolGEN is available as an open-source repository for contiguous development, at \url{https://github.com/tuoping/MolGEN}.

\section*{Acknowledgements}

P.T. acknowledges funding from the BIDMaP Postdoctoral Fellowship. P.T. acknowledges valued discussions with Dr. Peichen Zhong, Dr. Hao Tang and Dr. Chengbin Zhao. The authors acknowledge the resources of the National Energy Research Scientific Computing Center (NERSC), a Department of Energy Office of Science User Facility using NERSC award DOEERCAP0031751 'GenAI@NERSC'.

\bibliographystyle{unsrt}  

\bibliography{references}

\end{document}